\documentclass[doublecol]{epl2} 

\usepackage{graphicx}
\usepackage{dcolumn}
\usepackage{bm}
\usepackage{color}
\usepackage{amsmath,amssymb}

\newcommand{\ddx}[1]{\frac{\partial #1}{\partial x}}
\newcommand{\ddt}[1]{\frac{\partial #1}{\partial t}}
\newcommand{\ddxi}[1]{\frac{\partial #1}{\partial\xi}}

\newcommand{\lameppp}{\lambda_e^{\tiny{\substack{ ++ \\[-0.45mm] + }}}}

\renewcommand{\vec}[1]{\bm{#1}}

\graphicspath{{make_figures/fig_files/}}

\renewcommand{\revision}[1]{#1}

\title{Spirals and heteroclinic cycles in a spatially extended Rock--Paper--Scissors model of cyclic dominance}
\shorttitle{Spirals and heteroclinic cycles in a spatially extended RPS} 

\author{C. M. Postlethwaite\inst{1} \and A. M. Rucklidge\inst{2}}
\shortauthor{C. M. Postlethwaite \etal}

\institute{                    
  \inst{1} Department of Mathematics, Auckland University, Auckland, NZ\\
  \inst{2} School of Mathematics, University of Leeds, Leeds LS2 9JT, UK
}

\abstract{
Spatially extended versions of the cyclic-dominance Rock--Paper--Scissors model
have traveling wave (in one dimension) and spiral (in two dimensions) behavior.
The far field of the spirals behave like traveling waves, which themselves have
profiles reminiscent of heteroclinic cycles. We compute numerically a nonlinear
dispersion relation between the wavelength and wavespeed of the traveling
waves, and, together with insight from heteroclinic bifurcation theory and
further numerical results from 2D simulations, we are able to make predictions
about the overall structure and stability of spiral waves in 2D cyclic
dominance models.\\
\\
EPL, 117 (2017) 48006 DOI: https://doi.org/10.1209/0295-5075/117/48006}

\begin{document}

\maketitle

\section{Introduction}

Scissors cut Paper, Paper wraps Rock, Rock blunts Scissors: the simple game
of Rock--Paper--Scissors provides an appealing model for cyclic dominance
between competing populations or strategies in evolutionary game theory and
biology. The model has been invoked to explain the
repeated growth and decay of three competing strains of microbial
organisms~\cite{Kerr2002} and of three colour morphs of side-blotched
lizards~\cite{Sinervo1996}. 
In a well-mixed population, the dynamics of the model is dominated by
the presence of a heteroclinic cycle connecting the three equilibria 
where only one of the three species survives~\cite{May1975}. In continuum
models, non-zero initial populations can never lead to
extinction. However, in stochastic models, which include demographic 
fluctuations arising from the finite population size, fluctuations
will lead eventually to one species becoming extinct (say Rock). When this 
happens, Scissors no longer has any restraint on its population and so will
quickly wipe out Paper -- so fluctuations lead to one of the three
competitors eventually dominating~\cite{Kerr2006}, 

When spatial distribution and mobility of species is taken in to account, waves
of Rock can invade regions of Scissors, only to be invaded by Paper in turn; \revision{in a homogeous space,}
these waves can be organised into spirals, with roughly equal populations of
the three species at the core of each spiral, and each species dominating in
turn in the spiral arms~\cite{Reichenbach2007a}. \revision{Cyclic behaviour is also seen if spatial heterogeneity (patchiness) is also taken into account~\cite{schreiber2013}.}
As such, cyclic competition
with spatial structure has been invoked as a mechanism for explaining the
persistence of biodiversity in
nature~\cite{Reichenbach2007a,Reichenbach2008,Szolnoki2014}, and the
Rock--Paper--Scissors model with spatial structure is now an important
reference model for non-hierarchical competitive
relationships~\cite{Kerr2002,Szolnoki2014}.

The basic processes of growth and cyclic dominance between three species
can be modelled as~\cite{Frey2010}:

 \begin{equation}
 A+\phi \xrightarrow{1} A+A,\quad
 A+B \xrightarrow{\sigma} \phi+B,\quad
 A+B \xrightarrow{\zeta} B+B,
 \label{eq:RPS_processes}
 \end{equation}
where $A$ and $B$ are two of the three species and $\phi$ represents space for 
growth, with growth rate~$1$. Species~$B$ dominates $A$ either by removing 
it (at rate~$\sigma\geq0$) or by replacing it (at rate~$\zeta\geq0$). 
Processes for the other pairs of species are found by symmetry.
Individuals are placed on a spatial lattice and allowed to move to adjacent 
lattice sites. 
Mean field equations can be derived~\cite{Frey2010,Szczesny2014}:
 \begin{eqnarray}
 {\dot a} &= a (1 - \rho - (\sigma+\zeta)b + \zeta c) + \nabla^2 a, \nonumber\\
 {\dot b} &= b (1 - \rho - (\sigma+\zeta)c + \zeta a) + \nabla^2 b, 
 \label{eq:RPS_PDEs} \\
 {\dot c} &= c (1 - \rho - (\sigma+\zeta)a + \zeta b) + \nabla^2 c, \nonumber
 \end{eqnarray} 
where $(a,b,c)$ are non-negative functions of space $(x,y)$ and time~$t$,
representing the density of each of the three
species, and $\rho=a+b+c$. The coefficient of the diffusion terms is set to~$1$ by scaling $x$
and~$y$, and nonlinear diffusion effects~\cite{Szczesny2013} are suppressed.

Without diffusion, \eqref{eq:RPS_PDEs} has been well studied~\cite{May1975}. It
has five non-negative equilibria: the origin $(0,0,0)$, coexistence
$\frac{1}{3+\sigma}(1,1,1)$, and three on the coordinate axes, $(1,0,0)$,
$(0,1,0)$ and $(0,0,1)$. The origin is unstable; the coexistence point has
eigenvalues $-1$ and $\frac{1}{2}\left(\sigma\pm{i}\sqrt{3}(\sigma +
2\zeta)\right)/(3+\sigma)$, and the on-axis equilibria have eigenvalues $-1$,
$\zeta$ and $-(\sigma+\zeta)$. When $\sigma>0$, the coexistence point is
unstable and trajectories are attracted to a heteroclinic cycle between the
on-axis equilibria, approaching each in
turn, staying close for progressively longer times but never
stopping~\cite{May1975,Guckenheimer1988,Krupa1995}.

\begin{figure}
\setlength{\unitlength}{1mm}

\begin{center}
\begin{picture}(86,67)(0,0)

\put(7,30){\includegraphics[width=34mm]{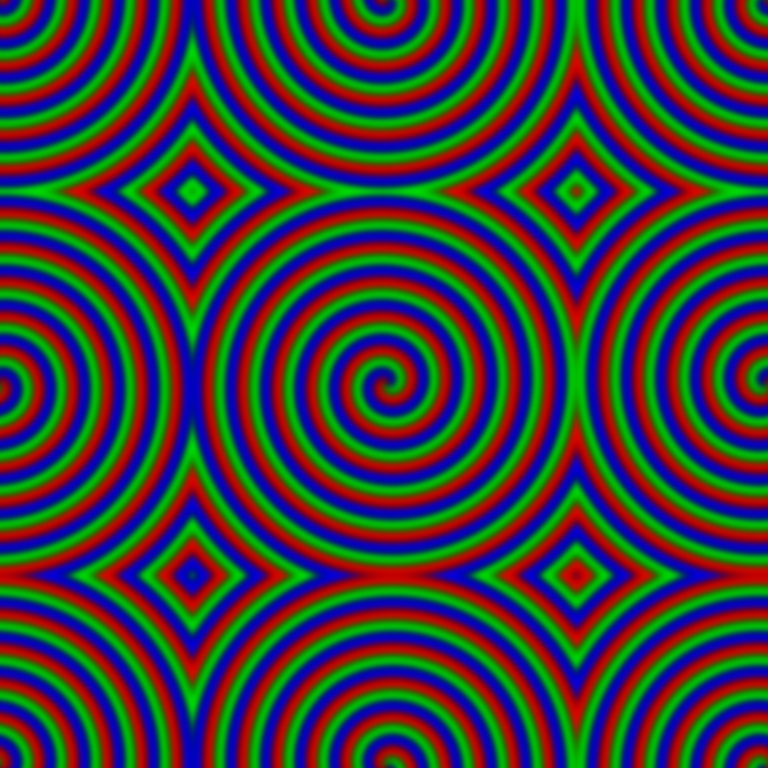}}

\put(4,2){\includegraphics[trim= 1.3cm 0cm 1.2cm 0cm,clip=true,width=38mm]{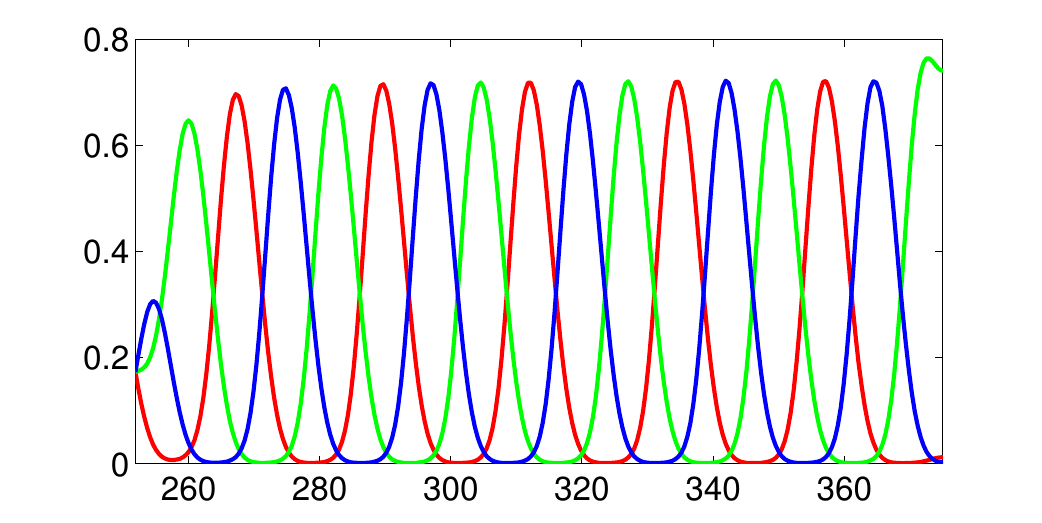}}

\put(48,46){\includegraphics[trim= 1.3cm 0cm 1.2cm 0cm,clip=true,width=38mm]{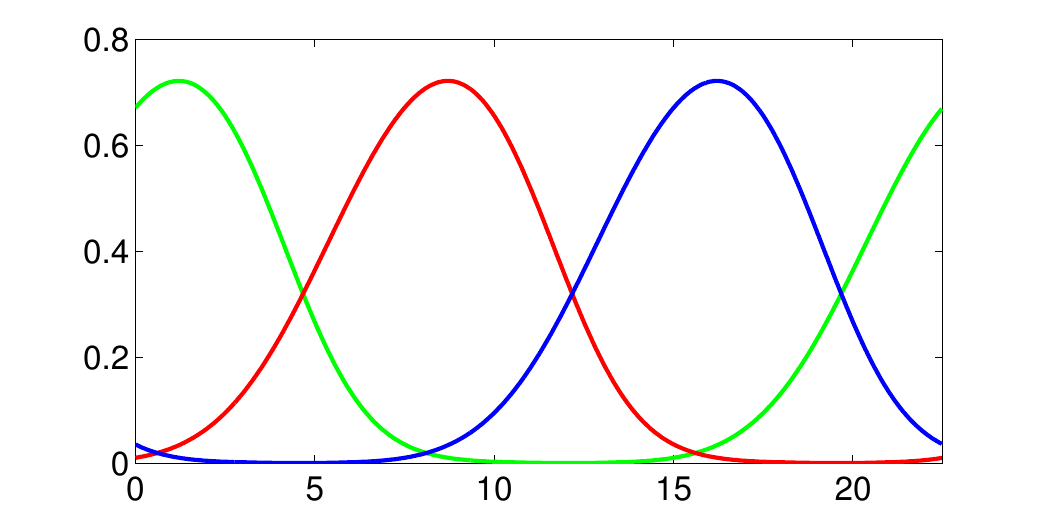}}
\put(48,24){\includegraphics[trim= 1.3cm 0cm 1.2cm 0cm,clip=true,width=38mm]{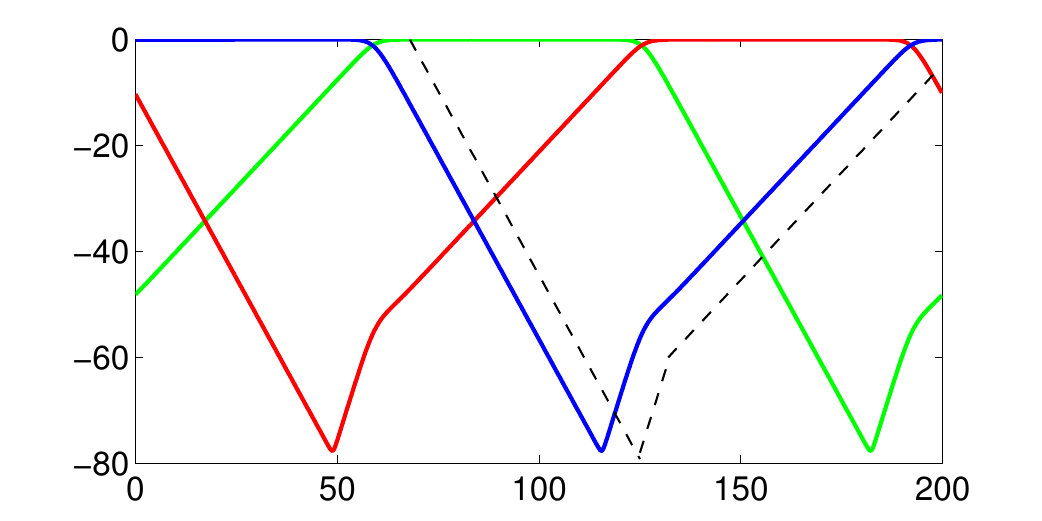}}
\put(48,2){\includegraphics[trim= 1.3cm 0cm 1.2cm 0cm,clip=true,width=38mm]{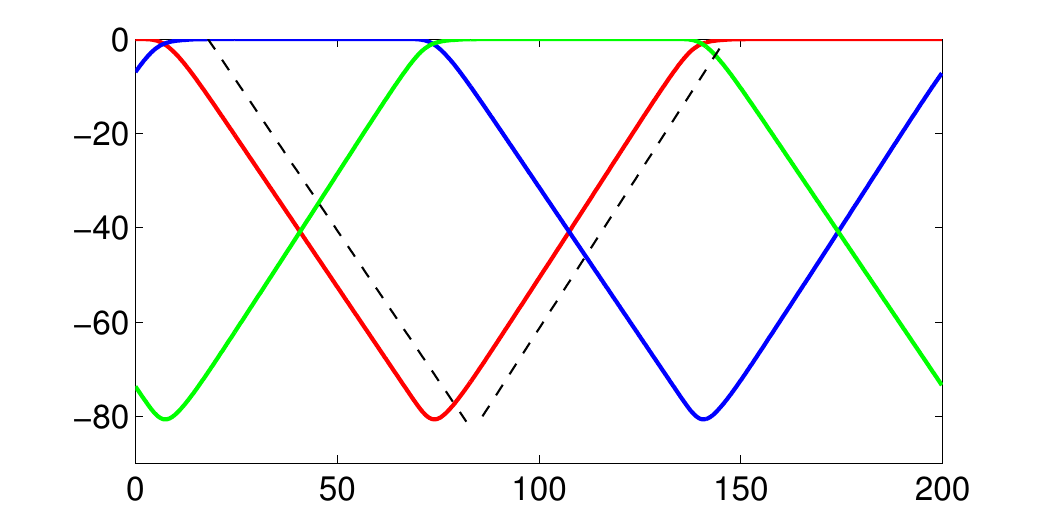}}

\put(0,65){(a)}
\put(44,65){(c)}
\put(44,43){(d)}
\put(0,21){(b)}
\put(44,21){(e)}
\put(83,0){{$x$}}
\put(39,0){{$x$}}

\put(0,10){{\rotatebox{90}{$a,b,c$}}}
\put(44,54){{\rotatebox{90}{$a,b,c$}}}
\put(44,28){{\rotatebox{90}{$\log a,b,c$}}}
\put(44,6){{\rotatebox{90}{$\log a,b,c$}}}

\put(69,34){\tiny{$\lambda_c^-$}}
\put(74,28){\tiny{$\lambda_c^+$}}
\put(79,35){\tiny{$\lambda_e^{++}$}}

\put(62,12){\tiny{$\lambda_c^-$}}
\put(67,7){\tiny{$\lambda_e^{+}$}}

\color{white}
\put(24,47){\line(1,0){8.5}}

\end{picture}
\end{center}

\vspace{-5ex}

 \caption{Numerical solutions of equations~\eqref{eq:RPS_PDEs}, with parameters $\sigma=3.2$,
$\zeta=0.8$ except in~(d,e); $a$, $b$ and $c$ are shown in red, green
and blue respectively. Panels~(a) and (b) show results from integration in 2D, with domain size $500\times500$; the spiral waves have
estimated clockwise rotation frequency $\Omega=0.440$ and far-field wavespeed $\gamma=1.576$
and wavelength $\Lambda=22.5$. Panel~(b) shows the profile along the white line
in~(a). Panels (c)--(e) show results from integrations in 1D. In
(c), the box size is  $\Lambda=22.5$ (c.f.\ the waves in (b)). Panel~(d) is
for a larger box ($\Lambda=200$), and $\zeta=0.2$; in log coordinates a
kink (change in slope) is evident in the upward phase of each curve. The estimated
wavespeed is $\gamma=1.059$. Panel~(e) has $\zeta=2$, and a profile without a kink. The estimated wavespeed is
$\gamma=2.834$. The dashed lines in (d) and~(e) show slopes as
indicated, labelled with eigenvalues from Table~\ref{tab:evals}.
\label{fig:TW}}
 \end{figure}
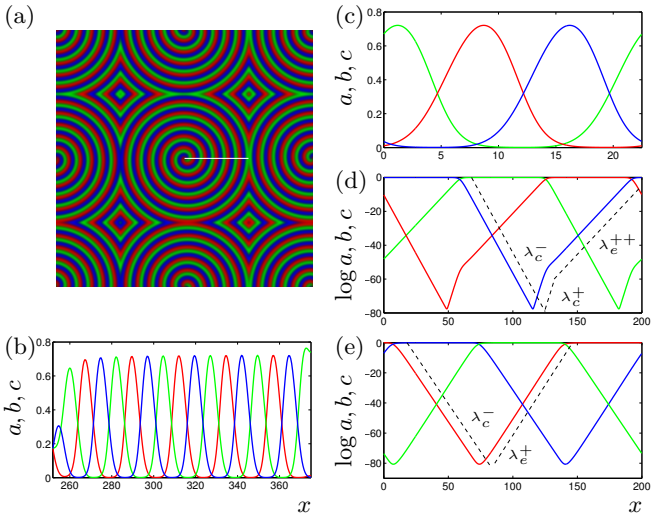

Numerical simulations of~\eqref{eq:RPS_PDEs} in sufficiently large two-dimensional
(2D) domains with periodic boundary conditions show a variety of behaviors as
parameters are changed~\cite{Frey2010,Orihashi2011}.
Stable
spiral patterns are readily found (Fig.~\ref{fig:TW}a), in which regions
dominated by~$A$ (red) are invaded by~$B$ (green), only to be invaded by~$C$
(blue). Comparing a cut through the core
(Fig.~\ref{fig:TW}b) with a one-dimensional (1D) solution with the same
wavelength (Fig.~\ref{fig:TW}c) demonstrates how the behavior far from the
core is essentially a 1D traveling wave (TW). Stable 1D TWs can be found
with arbitrarily long wavelength (Fig.~\ref{fig:TW}d,e), where (apart from
being periodic) the behavior closely resembles a heteroclinic cycle,
with traveling fronts between regions where one variable is close 
to~1 and the others are close to~0.

The question we ask is: can ideas from nonlinear
dynamics and heteroclinic cycles be used to analyze the properties (wavelength,
wavespeed and stability) of the 1D TWs and 2D spirals? Our approach is to consider the 1D TWs as periodic orbits in a moving
frame of reference, and use continuation techniques to calculate a nonlinear
relationship between the wavelength and wavespeed. 
We find parameter ranges in which these 1D TWs exist (between a Hopf bifurcation and three different types of heteroclinic bifurcation) and obtain partial information about their stability. The locations of the heteroclinic bifurcation are computed numerically, but in two of the three cases they coincide with straight-forward relations between eigenvalues.
We investigate 
2D solutions of the partial differential equations
(PDEs)~\eqref{eq:RPS_PDEs} over a range of parameter values, and 
show
numerically that the rotation frequency of the spiral
is related to the imaginary part of the eigenvalues of the coexistence fixed
point. Combining this information is enough to determine the overall
properties of the spiral.

\section{Analysis of traveling waves}

We first consider  equations~\eqref{eq:RPS_PDEs} in 1D,
and move to a right-traveling frame moving with wavespeed $\gamma>0$. We define
$\xi=x+\gamma{t}$, then $\ddx{}\rightarrow\ddxi{}$ and
$\ddt{}\rightarrow\gamma\ddxi{}+\ddt{}$.
Traveling wave solutions in the moving frame have $\ddt{}=0$, and so TW
solutions of~\eqref{eq:RPS_PDEs} correspond to periodic solutions of the
following set of six first-order ODEs:

 \begin{alignat}{3}
 {a}_{\xi} &=  u, \quad
 u_\xi &=  \gamma u - a(1-\rho-(\sigma+\zeta)b+\zeta c), \nonumber\\ 
 {b}_{\xi} &=  v, \quad
 v_\xi &=  \gamma v - b(1-\rho-(\sigma+\zeta)c+\zeta a), \label{eq:TW_ODEs} \\ 
 {c}_{\xi} &=  w, \quad
 w_\xi &=  \gamma w - c(1-\rho-(\sigma+\zeta)a+\zeta b).\nonumber
 \end{alignat} 
The period (in~$\xi$) of the
periodic solution corresponds to the wavelength~$\Lambda$ of the TW, and in numerical simulations of the PDEs in 1D with periodic boundary conditions, the size of the computational box.

Let
$\vec{x}=(a,u,b,v,c,w)$. The
coexistence and on-axis equilibria of the ODEs~\eqref{eq:TW_ODEs} are
$\vec{x}=\frac{1}{3+\sigma}(1,0,1,0,1,0)$, 
$(1,0,0,0,0,0)$, $(0,0,1,0,0,0)$ and
$(0,0,0,0,1,0)$. 
We label these equilibria $\xi_h$, $\xi_a$, $\xi_b$ and $\xi_c$
respectively.
\revision{The eigenvalues of the equilibrium $\xi_a$ are given in table~\ref{tab:evals}. By symmetry, $\xi_b$ and $\xi_c$ have the same eigenvalues.}
\revision{
It can easily be seen that} 
the four-dimensional subspace $\{c=w=0\}$  is invariant under the flow 
of~\eqref{eq:TW_ODEs}. Restricted to this
subspace, $\xi_a$ has a three-dimensional unstable manifold, and $\xi_b$ has a
two-dimensional stable manifold, which
generically intersect, and there is thus a robust heteroclinic connection
between $\xi_a$ and $\xi_b$. By symmetry, we have a robust heteroclinic cycle
between $\xi_a$, $\xi_b$ and $\xi_c$. 

\revision{
Following conventions used in the analysis of heteroclinic cycles (see e.g.~\cite{Krupa1995}) we label the eigenvalues as radial, contracting and expanding (see again table~\ref{tab:evals}). For $\xi_a$, the radial eigenvectors lie in the subspace $\{b=v=c=w=0\}$, the contracting eigenvectors in the subspace $\{b=v=0\}$ and the expanding eigenvectors in the subspace $\{c=w=0\}$. Note that this labelling does not exactly correspond with the definitions given in~\cite{Krupa1995} and other similar papers, mostly because of the presence of a positive contracting eigenvalue, which means that the unstable manifold of the equilibrium is not contained in the `expanding' subspace. However, we find the labelling useful because the eigenvalues play similar roles as to those seen in the literature, even though they do not exactly fit the definitions. }





\begin{table}
\caption{Eigenvalues of the on-axis equilibria 
 of~\eqref{eq:TW_ODEs}.
The radial and contracting eigenvalues are always real, and satisfy
$\lambda_r^-<0<\lambda_r^+$ and $\lambda_c^-<0<\lambda_c^+$. If $\gamma^2>4\zeta$, the
expanding eigenvalues are also real, and $\lambda_e^{++}>\lambda_e^+>0$. If
$\gamma^2<4\zeta$, the expanding eigenvalues $\lambda_e^R\pm{i}\lambda_e^I$ are complex,
and $\lambda_e^R>0$. 
\label{tab:evals}}
\begin{center}
\begin{tabular}{ll}
\strut
Label & Eigenvalues \\ \hline
\strut
Radial &  $\lambda_r^{\pm}=\frac{1}{2}\left(\gamma\pm\sqrt{\gamma^2+4}\right)$  \\
\hline
\strut
Contracting & $\lambda_c^{\pm}=\frac{1}{2}\left(\gamma\pm\sqrt{\gamma^2+4(\sigma+\zeta)}\right)$ \\
\hline
\strut
Expanding ($\gamma^2-4\zeta>0$) & $\lameppp=\frac{1}{2}\left(\gamma\pm\sqrt{\gamma^2-4\zeta}\right)$\\
\strut
Expanding ($\gamma^2-4\zeta<0$) & $\lambda_e^R\pm{i}\lambda_e^I=\frac{1}{2}\left(\gamma\pm{i}\sqrt{4\zeta-\gamma^2}\right)$
\end{tabular}
\end{center}
\end{table}

In numerical solutions of the PDEs~\eqref{eq:RPS_PDEs} in large 1D periodic domains of size~$\Lambda$, these
infinite-period heteroclinic cycles are excluded and we find
instead periodic solutions that lie close to the
heteroclinic cycle. These solutions spend a lot of ``time'' (a large interval 
in the $\xi$~variable) close to the
equilibria, where the components grow (or decay) exponentially with rates equal
to the relevant eigenvalues (see Fig.~\ref{fig:TW}d,e). In large domains
the TW profiles are thus determined by their wavelength~$\Lambda$ and
these eigenvalues. We find large-$\Lambda$ TWs with three
different profiles; two of which are shown in Fig.~\ref{fig:TW}(d,e). The
kinked profile in (d) takes the form:

 \begin{equation}
 \log a (\xi)=\left\{ \begin{matrix}
 0 & 0\leq \xi\leq \frac{\Lambda}{3} \\
 \lambda_c^-\left(\xi-\frac{\Lambda}{3}\right) & \frac{\Lambda}{3} < \xi \leq \frac{\Lambda}{3}+l \\
 \lambda_c^-l+\lambda_c^+\left(\xi-\frac{\Lambda}{3}-l\right) & \frac{\Lambda}{3}+l < \xi \leq \frac{2\Lambda}{3} \\
 \log a\left(\frac{2\Lambda}{3}\right)+\lambda_e^{++}\left(\xi-\frac{2\Lambda}{3}\right) & \frac{2\Lambda}{3} < \xi \leq \Lambda
 \end{matrix}
 \right. \nonumber
 \end{equation}
 and $b$ and $c$ are cyclic permutations, so $b(\xi)=a(\xi+
\frac{\Lambda}{3})$ and $c(\xi)=b(\xi+ \frac{\Lambda}{3})$.
The amount of ``decay'' in the contracting phase 
must match the amount of growth in the expanding phase, and these are both
of equal length. In this case, this means
there is a switch from decay to growth during the contracting phase at 
$\xi=\frac{\Lambda}{3}+l$, where
$l=\frac{\Lambda}{3}\frac{\lambda_c^++\lambda_e^{++}}{\lambda_c^+-\lambda_c^-}$ 
(and $0<l<\frac{\Lambda}{3}$),
and a change in the upwards slope (a kink) at $\xi=\frac{2\Lambda}{3}$.
The solution is continuous, periodic and $\log a(\Lambda)=0$. 
We have ignored the ``time'' taken for jumps between the equilibria (which
round the sharp corners of the profile) as these are short compared to
$\Lambda$, so long as $\Lambda$~is sufficiently large. Generically,
when the expanding eigenvalues are real,
we expect solutions leaving a neighbourhood of
an equilibrium to do so tangent to the leading expanding eigenvector: i.e. with an expansion rate equal to~$\lambda_e^+$. 
The profile observed in
Fig.~\ref{fig:TW}(d) is non-generic, and corresponds to an orbit flip, discussed further later.
The profile in Fig.~\ref{fig:TW}(e) has no kink,
and the rate of expansion is
$\lambda_e^+$ rather than $\lambda_e^{++}$. The third profile observed is
similar to that in Fig.~\ref{fig:TW}(d) except the expanding
eigenvalues are very slightly complex.

Although the heteroclinic cycle exists robustly in the ODEs, periodic solutions
cannot be found by forward integration since they are not
stable with respect to evolution in the $\xi$ variable. Instead, we identify a Hopf
bifurcation at the equilibrium $\xi_h$, and use the continuation software
AUTO~\cite{Doedel2001} to follow periodic orbits, treating the
wavespeed~$\gamma$ as a parameter, allowing the wavelength~$\Lambda$ to be
adjusted automatically.

\begin{figure}
\setlength{\unitlength}{1mm}
\begin{center}
\begin{picture}(86,50)(0,0)
\put(0,0){\includegraphics[trim= 1.9cm 0cm 2.2cm 0cm,clip=true,width=86mm]{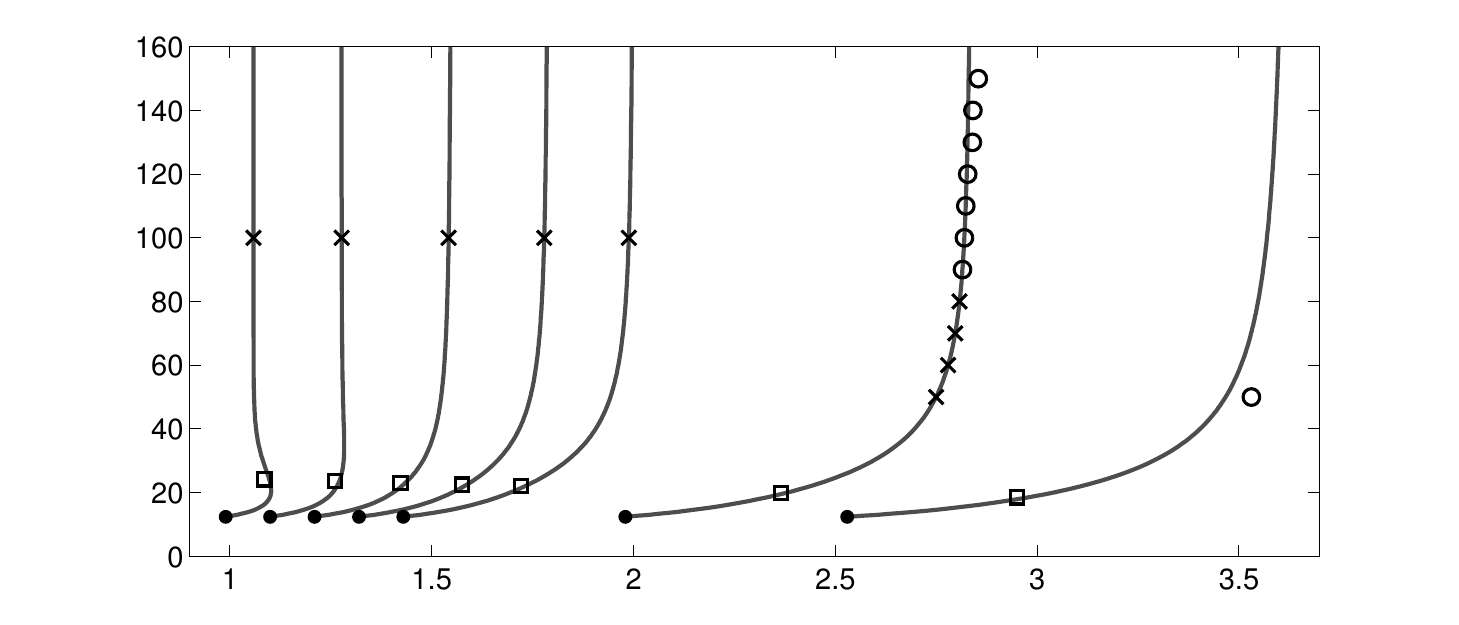}}
\put(84,2){{$\gamma$}}
\put(0,37.5){$\Lambda$}
\put(78,43){$\zeta=3$}
\put(57,43){$\zeta=2$}
\put(33,43){$\zeta=1$}
\put(26,47){$\zeta=0.8$}
\put(20,43){$\zeta=0.6$}
\put(12,47){$\zeta=0.4$}
\put(5,43){$\zeta=0.2$}
\end{picture}
\end{center}


\caption{The wavelength (period in~$\xi$)
$\Lambda$, as $\gamma$ is varied, of periodic orbits in the
ODEs~\eqref{eq:TW_ODEs}, computed using AUTO, with $\sigma=3.2$ and
values of $\zeta$ as indicated. Each curve of periodic
orbits arises in a Hopf bifurcation on the left (black dot), and
ends in a heteroclinic (long-period) bifurcation on the right. Effectively
these curves are nonlinear dispersion relations for TWs in the
PDEs. Symbols indicate the results of 1D TW and 2D spiral solutions of the
PDEs~\eqref{eq:RPS_PDEs}, as described in the text.
\label{fig:per_orbs}}
\end{figure}
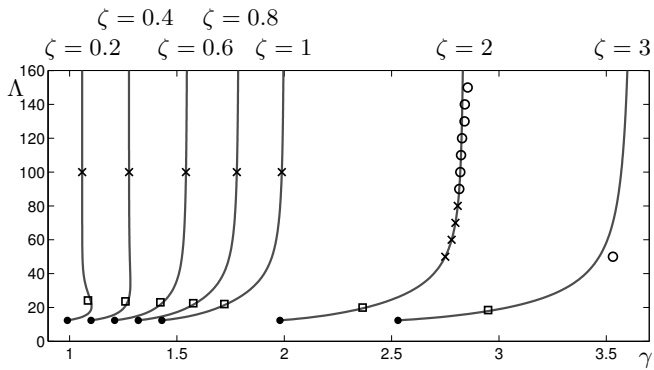

The Jacobian matrix at $\xi_h$ has pure imaginary eigenvalues~$\pm i\omega_H$ when
$\gamma=\gamma_H(\sigma,\zeta)$, where

 \begin{equation}\label{eq:hopf}
 \gamma_H(\sigma,\zeta)\equiv\frac{\sqrt{3}(\sigma+2\zeta)}{\sqrt{2\sigma(\sigma+3)}},
 \quad\text{and}\quad
 \omega_H^2=\frac{\sigma}{2(\sigma+3)},
 \end{equation}
at which point a Hopf bifurcation creates periodic orbits of
period~$\Lambda_H=\frac{2\pi}{\omega_H}$. Fig.~\ref{fig:per_orbs} shows, for
$\sigma=3.2$ and a range of values of~$\zeta$, the wavelength
(period in~$\xi$)~$\Lambda$ as $\gamma$ is varied. The range of~$\gamma$ for
which periodic solutions can be found depends on $\sigma$ and~$\zeta$; each
branch starts at $\gamma_H$ and terminates with infinite~$\Lambda$ in a
heteroclinic bifurcation.

\begin{figure}
\setlength{\unitlength}{1mm}
\begin{center}
\begin{picture}(86,75)(0,0)
\put(0,0){\includegraphics[trim= 1.3cm 0cm 1.5cm 0.9cm,clip=true,width=86mm]{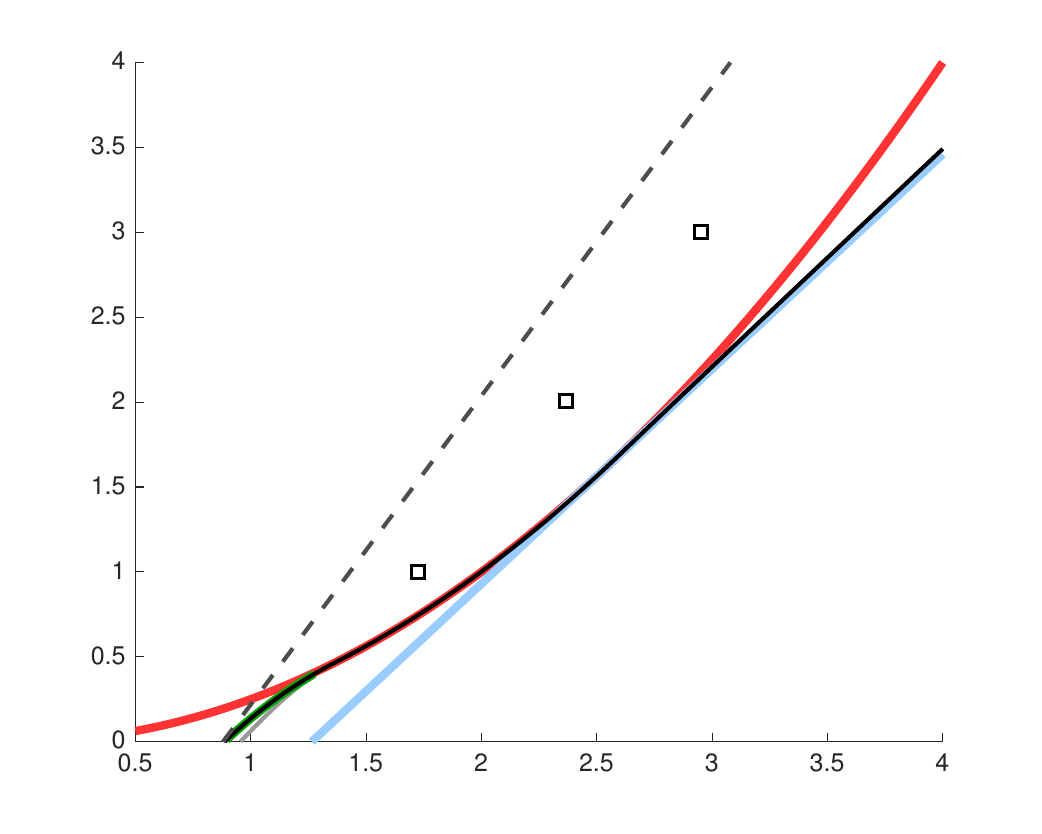}}                             
\put(80,3){{$\gamma$}}
\put(0,71){$\zeta$}

\put(8,37){\includegraphics[trim= 0cm 0cm 0cm 0cm,clip=true,width=18mm]{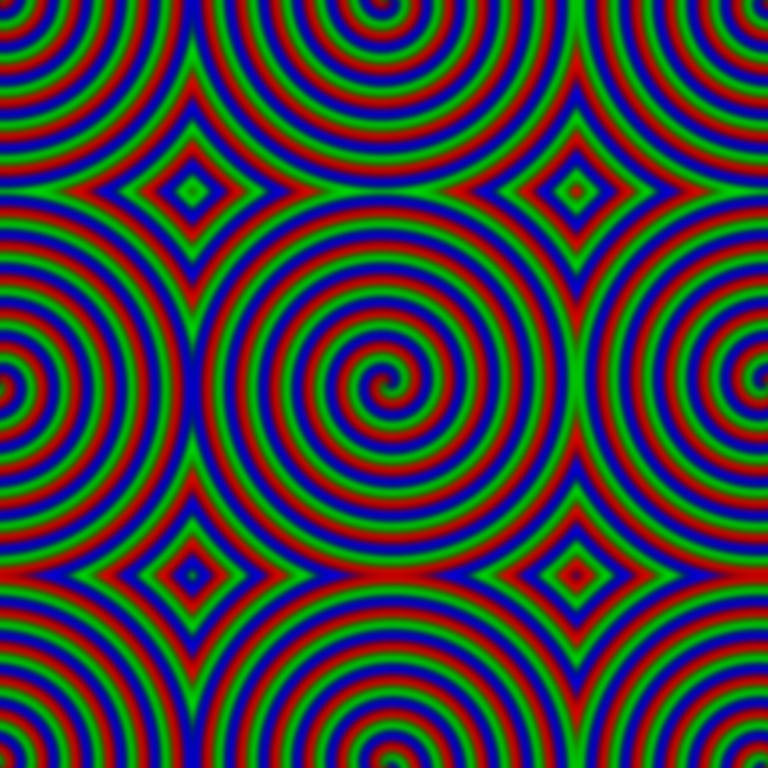}}
\put(26,37){\vector(3,-4){7}}
\put(8,57){\includegraphics[trim= 0cm 0cm 0cm 0cm,clip=true,width=18mm]{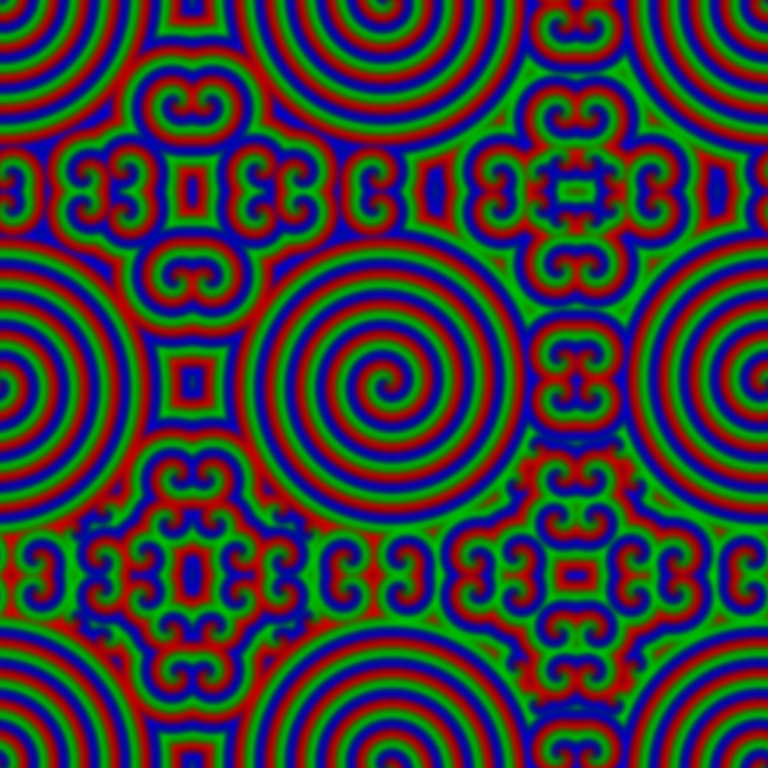}}
\put(26,57){\vector(3,-2){20}}
\put(28,57){\includegraphics[trim= 0cm 0cm 0cm 0cm,clip=true,width=18mm]{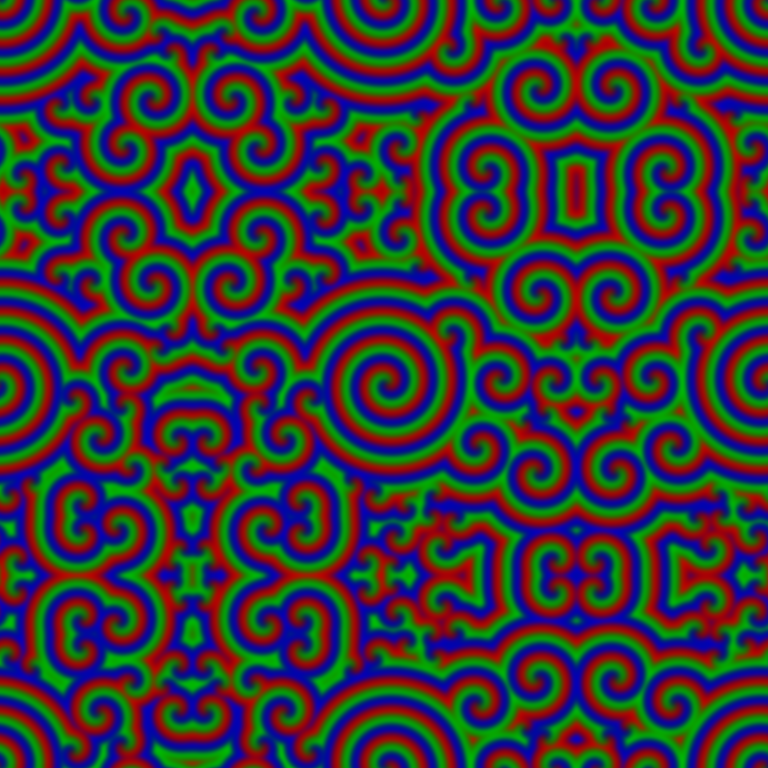}}
\put(46,59){\vector(1,0){13}}
\put(14,15){\rotatebox{53}{Hopf: $\gamma=\gamma_H$}}

\put(50,25){\textcolor{blue}{$1$}}
\put(80,65){\textcolor{blue}{$2$}}
\put(53,44){\textcolor{blue}{$3$}}
\put(23,12){\textcolor{blue}{$4$}}

\put(64,70){$\lambda_e^+=\lambda_e^{++}$}
\put(71,69){\vector(1,-1){4}}

\put(71,50){$|\lambda_c^-|=\lambda_e^+$}
\put(78,53){\vector(0,1){5}}

\put(34,13){$|\lambda_c^-|=\lambda_e^{++}$}
\put(40,16){\vector(-1,1){4}}

\put(58,16){\includegraphics[trim= 0.cm 0cm 0.1cm 0cm,clip=true,width=28mm]{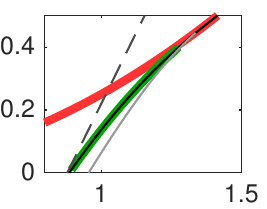}}

\put(74,20){SN}
\put(73,21){\vector(-1,0){4}}

\end{picture}
\end{center}

\vspace{-6ex}

\caption{Bifurcation diagram for the
ODEs~\eqref{eq:TW_ODEs}, in $(\gamma,\zeta)$ parameter space, with $\sigma=3.2$.
The blue line ($\zeta=\sqrt{\frac{\sigma}{2}}\gamma-\frac{\sigma}{2}$)
and red curve ($4\zeta=\gamma^2$) are tangent at $(\gamma, \zeta)=(\sqrt{2\sigma},\sigma/2)$ and divide the parameter space into four
regions, labeled by blue numbers, and defined in table~\ref{tab:pspace}. The green curve is the locus of a heteroclinic orbit flip.
The dark grey \revision{dashed} line is a curve of Hopf bifurcations. Periodic orbits bifurcate to the
right of this line and disappear in a curve of heteroclinic bifurcations
(black). A curve of saddle-node bifurcations of periodic orbits (light
grey) exists for smaller $\zeta$. The upper insets show 2D simulations at
the indicated parameter values.
The lower inset is a zoom near the saddle-node (SN) and orbit flip (green) bifurcations.
\label{fig:pspace}} 
\end{figure}
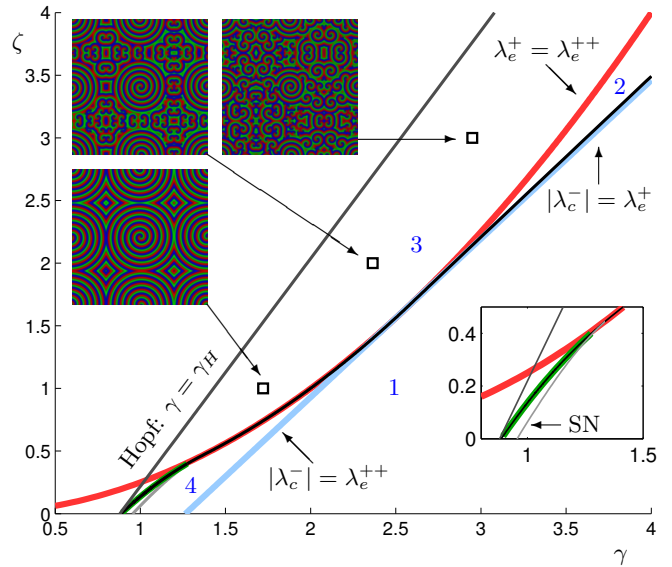

\revision{
In Fig.~\ref{fig:pspace} we show a bifurcation diagram of the
ODEs~\eqref{eq:TW_ODEs} (computed by AUTO)
in $(\gamma,\zeta)$ space. The red and blue curves correspond
to simple equalities of the eigenvalues, as indicated in the figure, and divide
the parameter space into four labelled regions, defined in
table~\ref{tab:pspace}. 
Periodic solutions bifurcate to the right of the
Hopf bifurcation, given by~\eqref{eq:hopf}, into region 3 (except for very small $\zeta$)
 and disappear in the
heteroclinic bifurcation curve (black) on the right. In 1D PDE simulations, this corresponds to observing small wavelength travelling waves just after the Hopf bifurcation (in region 3) which grow in wavelength as $\gamma$  increases and disappear at the black curve. Note that the dynamics for the PDEs~\eqref{eq:RPS_PDEs} and the ODEs~\eqref{eq:TW_ODEs} only coincide when the travelling wave solutions exist, i.e. betwen the Hopf curve (dashed line) and the heteroclinic curve (black curve).} 

\begin{largetable}
\caption{Definitions of the regions of parameter space shown
in Fig.~\ref{fig:pspace} and eigenvalue properties therein.
 \label{tab:pspace}}
\begin{center}
\begin{tabular}{ccc}
Region & Definition & Eigenvalue properties \\ \hline
1 &  $\zeta<\sqrt{\frac{\sigma}{2}}\gamma-\frac{\sigma}{2}$ & $\lameppp\in\mathbb{R}$, $\lambda_e^+<|\lambda_c^-|<\lambda_e^{++}$  \\
\hline
2 & $\zeta>\frac{\sigma}{2}$, $ \sqrt{\frac{\sigma}{2}}\gamma-\frac{\sigma}{2}<\zeta<\frac{\gamma^2}{4}$ &
 $\lameppp\in\mathbb{R}$, $|\lambda_c^-|<\lambda_e^+<\lambda_e^{++}$ \\
\hline
3 & $\zeta>\frac{\gamma^2}{4}$ & $\lameppp\in\mathbb{C}$ \\
\hline
4 & $\zeta<\frac{\sigma}{2}$, $ \sqrt{\frac{\sigma}{2}}\gamma-\frac{\sigma}{2}<\zeta<\frac{\gamma^2}{4}$ & $\lameppp\in\mathbb{R}$, $\lambda_e^+<\lambda_e^{++}<|\lambda_c^-|$
\end{tabular}
\end{center}
\end{largetable}

We observe from the numerical results that the heteroclinic bifurcation in Fig.~\ref{fig:pspace}
shows three different behaviors, overlying the green, red and blue curves in 
different parameter regimes,
corresponding to the three large-$\Lambda$ TW profiles discussed earlier.
Note that heteroclinic bifurcations cannot occur in the interiors of regions 2 or~3. 
In region~2, a large-$\Lambda$ TW profile would require  $l>\frac{\Lambda}{3}$, which cannot occur. 
In region~3, the expanding eigenvalues are complex. 
In the large~$\Lambda$ limit, complex eigenvalues are excluded: the invariance
of the subspace $\{a=u=0\}$ means that $a$ cannot change sign along
trajectories.

When $\zeta>\frac{\sigma}{2}=1.6$, the
heteroclinic bifurcation occurs on the blue curve, along which the negative
contracting and leading expanding eigenvalues are equal in magnitude, and the
TW has an unkinked profile (Fig.~\ref{fig:TW}(e)). This is a
heteroclinic resonance bifurcation~\cite{Scheel1992}. For
$0.4<\zeta<\frac{\sigma}{2}=1.6$, the heteroclinic bifurcation occurs on the
red curve, along which the expanding eigenvalues switch from complex
to real (a variant of a Belyakov--Devaney bifurcation~\cite{Champneys1998}), and
the TW has a kinked profile. For
$0<\zeta<0.4$, the periodic orbit undergoes a saddle-node bifurcation before
the heteroclinic bifurcation; the fold can be seen in the curve for $\zeta=0.2$
in Fig.~\ref{fig:per_orbs}. Here, the heteroclinic
bifurcation coincides with an orbit flip
bifurcation~\cite{Homburg2000}, indicated in green in Fig.~\ref{fig:pspace}. 
The TW has a kinked profile,  as in Fig.~\ref{fig:TW}(d). 
The location of the orbit flip is computed by solving a boundary value problem in the four-dimensional invariant subspace $\{c=w=0\}$ that requires that the heteroclinic solutions is tangent to the $\lambda_e^{++}$ eigenvector.

Returning to the PDEs~\eqref{eq:RPS_PDEs}, we computed solutions
over a range of values of~$\sigma$, $\zeta$ and domain size. We imposed periodic boundary conditions, and
used fast Fourier transforms and second-order exponential time differencing~\cite{Cox2002}. 
In 2D, we mainly used $1000\times1000$ domains, with
$1536$ Fourier modes in each direction.
We estimated
speeds of TWs (in 1D) and rotation frequencies and
far-field wavelengths and wavespeeds of spirals (in 2D).

In 1D, with $\sigma=3.2$ and $\zeta<\frac{\sigma}{2}=1.6$, we are able to
find stable TWs for all box sizes larger than $\Lambda_H$. For
$\zeta>\frac{\sigma}{2}$, we find that TWs are stable in smaller
boxes, and unstable in larger boxes, with a decreasing range of stable boxes sizes as
$\zeta$~is increased. For $\zeta=3$, we are unable to find any stable 
TWs. The crosses (resp.\ open circles) in
Fig.~\ref{fig:per_orbs} show the observed wavespeeds of stable (resp.\
unstable) TWs for a range of $\zeta$ and box sizes. 
In this context, by ``stable'' we are referring to how the TWs
evolves in time with a fixed wavelength. A full treatment of stabiliity 
would include convective and absolute instability of the~TWs.

In 2D, spiral waves (or more complex solutions) are usually found if the domain
is large enough. We use initial conditions that are one half $a$ and a quarter
each $b$ and~$c$, as in~\cite{Szczesny2013}.
When we find spirals,
we locate the core (where $a=b=c$) and compute the far-field
wavelength by taking a cut through the core
(Fig.~\ref{fig:TW}(a,b)). The angular frequency~$\Omega$ is obtained
from a timeseries (the temporal period is $2\pi/\Omega$), and the wavespeed is 
$\gamma=\Lambda\Omega/2\pi$. For
$\sigma=3.2$ and a selection of~$\zeta$, we have included in
Fig.~\ref{fig:pspace} three examples, along with their
$(\gamma,\zeta)$ values, and in Fig.~\ref{fig:per_orbs} (as open squares) the
$(\gamma,\Lambda)$ values estimated from spiral solutions. The fact that the
open square symbols lie on the continuation curves from AUTO confirms that the
far field of the spirals 
obeys the same nonlinear dispersion relation as 1D solutions.

\begin{figure}
\setlength{\unitlength}{1mm}
\begin{center}
\begin{picture}(86,41)(0,0)
\put(0,0){\includegraphics[trim= 1.9cm 0.2cm 2.1cm 0.5cm,clip=true,width=86mm]{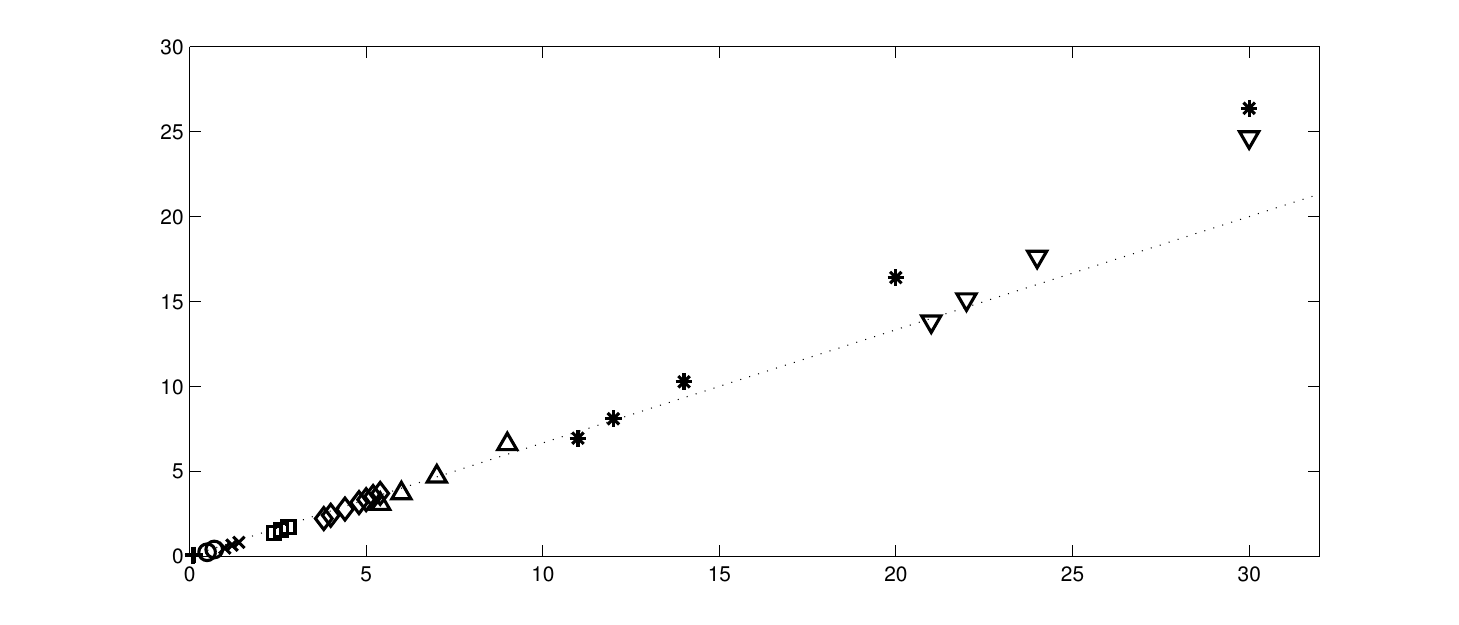}}
\put(69,0){{$\sigma+2\zeta$}}
\put(-2,15){\rotatebox{90}{$2\Omega(\sigma+3)/\sqrt{3}$}}

\put(6,19){\includegraphics[trim= 1.9cm 0cm 2.1cm 0.2cm,clip=true,width=41mm]{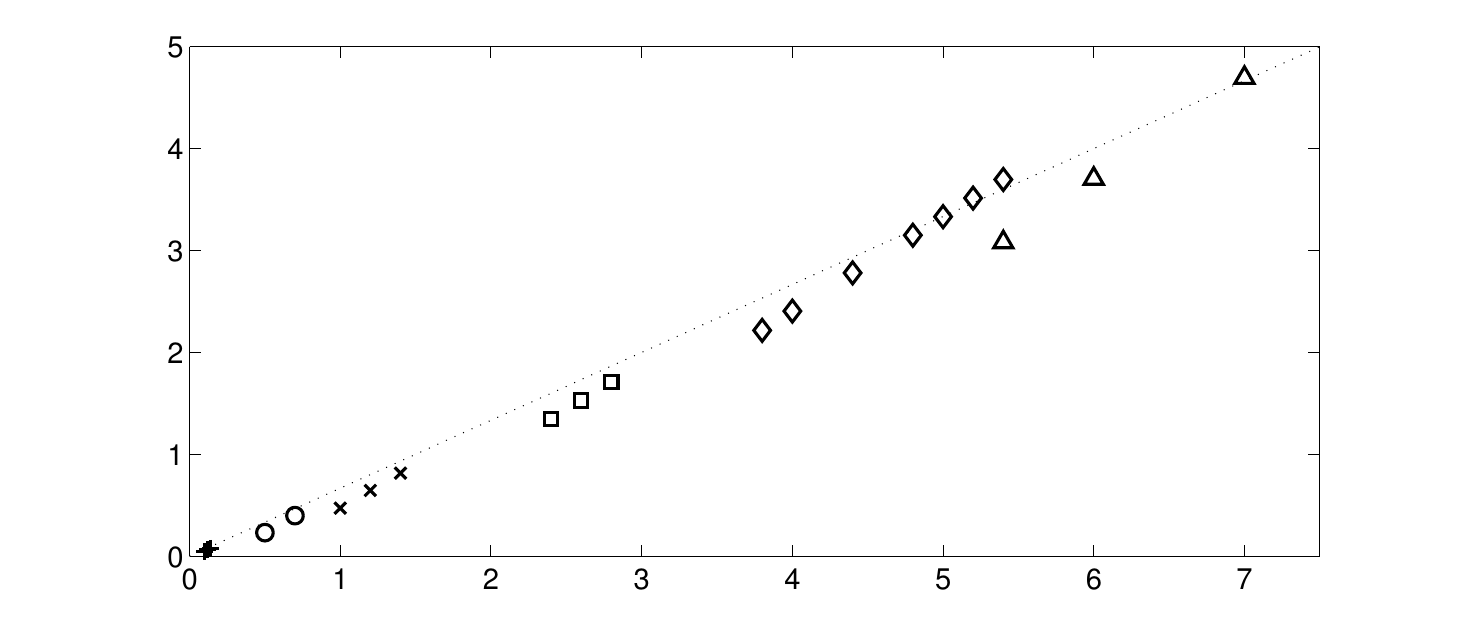}}

\end{picture}
\end{center}


\caption{The scaled spiral frequency
$2\Omega(\sigma+3)/\sqrt{3}$ plotted against $\sigma+2\zeta$, for results from
2D simulations over a range of $\sigma$ and~$\zeta$. The dotted line has a
slope of $\frac{2}{3}$. The inset shows a zoom of the origin. Different symbols
correspond to different values of $\sigma$: $(0.1, 0.5, 1, 2, 3.2, 5, 10, 20) =
(+,\bigcirc,\times,\square,\Diamond,\bigtriangleup,\star,\bigtriangledown)$.
\label{fig:4}}

\end{figure}

We now have two relations between three quantities, the rotation
frequency~$\Omega$ of the 2D spiral, and the wavespeed~$\gamma$ and
wavelength~$\Lambda$ of the 1D TWs in the far field.
When locating the
core 
we observed that the common value of the
three variables is almost $\frac{1}{3+\sigma}$, the value from the coexistence
equilibrium. 
We therefore compared the rotation frequency~$\Omega$ to the
imaginary part of the complex eigenvalue at the coexistence equilibrium,
plotting (in Fig.~\ref{fig:4})
$\frac{2}{\sqrt{3}}\Omega(\sigma+3)$ against $\sigma+2\zeta$.
The data almost
collapses on to a straight line of slope (approximately)~$\frac{2}{3}$, over
the range of~$\sigma$ and $\zeta$ that we investigated. If
$\Omega$ were \emph{equal} to the imaginary part of the complex eigenvalue, the
slope would be~$1$.
For each value of~$\sigma$, indicated by the symbols in Fig.~\ref{fig:4}, the
value ``$\frac{2}{3}$'' depends weakly on~$\zeta$, with increasing departure
from this value for larger~$\sigma$.

This data collapse is sufficient to give a complete prediction for the
properties of a spiral: the angular frequency~$\Omega$ is set by the core and
is approximately $\frac{2}{3}\frac{\sqrt{3}}{2}(\sigma+2\zeta)/(3+\sigma)$. The
other two quantities $\gamma$ and $\Lambda$ are set by
$\gamma=\Lambda\Omega/2\pi$ and the nonlinear dispersion relation in
Fig.~\ref{fig:per_orbs}.

It remains to consider the far-field stability of the spirals. 
\revision{As can be seen in the insets in figure~\ref{fig:pspace}, the size of the spirals in the 2D simulations appears to decrease as $\zeta$ is increased. }
With
$\sigma=3.2$, we find $1000\times1000$ domain-filling 2D spirals (as in
Fig.~\ref{fig:TW}a) over the range $0.2\leq\zeta\leq1.2$. For values of~$\zeta$
outside this range, the far field of the spiral breaks up, 
and for $\zeta=0.2$ and $\zeta\geq1.1$, this
is also seen in a larger domains. This pattern is repeated with
other values of~$\sigma$: in the range $0.1\leq\sigma\leq20$, 
we find stable domain-filling spirals 
in a finite range of~$\zeta$; for small~$\sigma$ and~$\zeta$, 
the wavelengths of the
spirals 
are so big that only a few turns fit in to
the domain. 
The spiral wavelengths are typically about $2\Lambda_H$, which suggests
from~\eqref{eq:hopf} that 
for small~$\sigma$ the wavelength scales as~$\sigma^{-\frac{1}{2}}$.
The same scaling can be deduced from the results (based on a completely
different approach) of~\cite{Szczesny2014}

\section{Discussion}

Models related to~\eqref{eq:RPS_PDEs} with one species ($b=c=0$, the
Fisher--KPP equation) and with two species ($c=0$, the Lotka--Volterra system)
are used to describe moving fronts between regions of different genes or
species. Although in these models the equilibria having real eigenvalues
imposes a constraint on the wavespeed, the speed that is observed is set by
details of the initial population profiles. In the case of the Fisher--KPP
equation, there is a lower bound of~$2$ on the front propagation
speed~\cite{Fisher1937}. Our success in describing the dynamics of spirals in
the three-species case, without reference to details of the initial conditions,
relies on the interesting structures being periodic TW, rather than fronts, and
on these TW arising in a Hopf bifurcation, which is absent in the Fisher--KPP
equantion and the Lotka--Volterra system.

Our approach complements that taken by~\cite{Reichenbach2008}, where spirals
are described in terms of a Complex Ginzburg--Landau equation (CGLE). Strictly,
this description requires a Hopf bifurcation from the coexistence equilibrium
in~\eqref{eq:RPS_PDEs}. There is a (degenerate) Hopf bifurcation at $\sigma=0$.
Its degeneracy can be broken by including the effect of
mutation~\cite{Mobilia2010}, and an asymptotic description of small-amplitude
(weakly nonlinear) spirals close to the coexistence equilibrium can be inferred
by reducing~\eqref{eq:RPS_PDEs} (with mutation) to the
CGLE~\cite{Szczesny2013,Szczesny2014}. In contrast, our approach treats the TW
as fully nonlinear, close to a heteroclinic cycle. The stability predictions
cannot be compared directly, and true 2D spirals are in between these two
extremes, but both approaches yield a $\sigma^{-\frac{1}{2}}$ scaling (for
small~$\sigma$) of the wavelength of the~TWs.

In spite of the prevalence of spirals in this model, spirals have yet to be
observed in nature or in experiments involving non-hierarchical competitive
relationships between species. It may be that the model is too simple and
neglects important effects~\cite{Weber2014,Kelsic2015}, it may be that the
system is operating in a regime where spirals are entirely fragmented (and 
indeed the parameters are hard to estimate~\cite{Warne2016}), or it may be that
the spirals that should be present are in fact larger than the domain under
consideration or smaller than the spacing between sampling
locations~\cite{Mobilia2016}. Notwithstanding these caveats, the
Rock--Paper--Scissors model remains an appealing reference model for cyclic
competition.

\acknowledgments
The authors would like to thank Graham Donovan, Edgar Knobloch, 
Mauro Mobilia and Hinke Osinga
for helpful discussions regarding this work. This work was started 
during a visit of CMP to Leeds partly funded by a Scheme~2 grant 
from the London Mathematical Society and the University of Auckland.

\bibliographystyle{eplbib}
\bibliography{allrefs}

\end{document}